\RequirePackage{lineno}
\documentclass[prl,twocolumn,nofootinbib,floatfix,superscriptaddress]{revtex4-1}

\usepackage{cancel}
\usepackage{graphicx}
\usepackage{epsfig}
\usepackage{amsmath,amsfonts,amssymb}
\usepackage{t1enc}
\usepackage{bm}

\newcommand{\TeV}{{\rm TeV}}

\begin{document}

%\linenumbers

\title{A closer look at the possible CMS signal of a new gauge boson}

\author{J. A. Aguilar-Saavedra}
\affiliation{Departamento de F\'{\i}sica Te\'orica y del Cosmos, Universidad de Granada,
 Granada, Spain}
 \affiliation{PH-TH Department, CERN, CH-1211 Geneva 23, Switzerland}
\author{F. R. Joaquim}
\affiliation{Departamento de F\'{\i}sica and CFTP, Instituto Superior T\'ecnico, Universidade de Lisboa, Lisboa, Portugal}

%\begin{document}

\begin{abstract}
The CMS collaboration has recently reported a 2.8$\sigma$ excess of $eejj$ events with an invariant mass around 2 TeV. This observation can be explained in the context of standard model extensions with new gauge bosons $W'$, $Z'$ and heavy neutrinos coupling (mainly) to the electron. We discuss additional signals that allow to confirm or discard the $W'$ and $Z'$ hypotheses.

\end{abstract}

\maketitle

The search for physics beyond the standard model (SM) is one of the flagship programs of the CERN Large Hadron Collider (LHC). Among the various new physics scenarios leading to the SM as a low-energy limit, models that incorporate a mechanism to generate light neutrino masses ---in particular, a seesaw mechanism--- are of special interest, since neutrino masses require physics beyond the SM. The new degrees of freedom involved in neutrino mass generation may also play a crucial role in the explanation of the observed baryon asymmetry of the Universe through the mechanism of leptogenesis (see~\cite{Branco:2011zb} for a review), and also enhance the rates of SM-suppressed lepton flavour violating decays like $\mu \to e\gamma$~\cite{Abada:2007ux}. Therefore,  potential LHC discoveries in this direction may have wide implications on physics at the high intensity frontier as well as on cosmology.

Recently, the CMS collaboration has reported a $2.8\sigma$ excess~\cite{Khachatryan:2014dka} in the search of heavy neutrinos mediating a type-I seesaw~\cite{Minkowski:1977sc} in the context of left-right (LR) symmetric models~\cite{Pati:1974yy}. 
One of the main features of these models is the existence of new heavy gauge bosons $W'$, $Z'$ (also often denoted as $W_R$, $Z_R$), which are the right-handed (RH) counterparts of the SM left-handed (LH) $W^\pm$ and $W^3$ bosons. At the same time, new heavy neutrinos naturally fit in SU(2)$_R$ doublets together with the RH charged lepton fields, being their charged interactions described by 
\begin{eqnarray}
\mathcal{L} & = & -\frac{g'}{\sqrt 2} V_{\ell N}^R \bar \ell_R \gamma^\mu N_{R} W_{\mu}^{\prime-} \notag \\
& &  -\frac{g}{\sqrt 2} V_{\ell N}^L \bar \ell_L \gamma^\mu N_{L} W_{\mu}^{-} + \text{H.c.} \,.
\label{ec:cc}
\end{eqnarray}
Here, $g,g'$ are the $\text{SU}(2)_{L,R}$ gauge couplings, respectively; $\ell = e,\mu,\tau$ and $V_{\ell N}^{L,R}$ denote the $\ell N$ mixing parameters in the LH and RH sectors. The RH mixings are of order unity, satisfying $|V_{eN}^R|^2+|V_{\mu N}^R|^2+|V_{\tau N}^R|^2=1$. On the other hand, $V_{\ell N}^L$ are seesaw-suppressed. The LH mixing effects in neutrino decays are subleading except when the $W'$ boson is much heavier than the neutrino, $M_{W'} \gg m_N$~\cite{delAguila:2009bb}. Consequently, RH interactions determine the production and decay modes of the heavy neutrino.
The particular process targeted by the CMS search is 
\begin{equation}
pp \to W' \to \ell_i N \to \ell_i\ell_k W^{\prime *} \to \ell_i \ell_k  jj \,,
\label{ec:Wprod}
\end{equation}
which can take place if $m_N < M_{W'}$~\cite{Keung:1983uu}. The produced heavy neutrinos undergo a three-body decay $N \to \ell_k q \bar q'$, with $q \bar q' = u\bar d, c\bar s,t \bar b$ leading ---among other signatures--- to a final state with two charged leptons and two jets, with an invariant mass $m_{\ell_i \ell_k jj} \sim M_{W'}$. The CMS Collaboration reports a $2.8\sigma$ excess in the dielectron channel ($\ell_i \ell_k = ee$), at $m_{eejj} \sim 2~\text{TeV}$, with almost all events having opposite-sign leptons, and no excess in the dimuon channel. (The $e\mu$ and $\tau$ lepton channels have not been analyzed.) The cross section of the excess can be trivially fitted with a heavy neutrino coupling to the electron $V_{eN} \simeq 1$ and $g'/g \sim 0.6$~\cite{Deppisch:2014qpa}, which also helps evading limits on $W'$ production in dijet and $t \bar b$ final states~\cite{Heikinheimo:2014tba} and relaxes indirect limits on the $W'$ mass~\cite{Barenboim:1996nd}. However, the CMS Collaboration claims that the kinematics of the excess ---here implicitly assuming that $N$ couples only to the electron--- is apparently not consistent with~(\ref{ec:Wprod}) since no localized excess in other unspecified distributions are found. 

In this Letter we address the interpretation of this excess in terms of process~(\ref{ec:Wprod}) but with a general flavor structure. In particular, we allow a sizable mixing $V_{\tau N}^R$, which smears the kinematical distributions as a fraction of the dielectron events result from $\ell_i \ell_k = e\tau,\tau e$ with subsequent $\tau$ decay. In addition, we consider the excess in the context of heavy neutrino pair production mediated by a new neutral boson $Z'$,\begin{equation}
pp \to Z^\prime \to N N \to \ell_i W \ell_k W\,, \quad W \to jj\,,
\label{ec:Zprod}
\end{equation}
which has a different kinematics with two more jets in the final state. This signal has already been considered in another context~\cite{Dobrescu:2014esa}.

Let us first discuss, based on available experimental data, the possible new physics sources of the CMS $eejj$ excess. The non-observation of a $\mu \mu jj$ excess hints to a non-universal flavor structure, with the presence of a new particle that mainly couples to the electron, and negligible coupling to the muon. A sizable coupling to the $\tau$ lepton is also acceptable but cannot be dominant, otherwise a signal would appear with similar strength in dielectron and dimuon final states from $\tau$ decays. If this new particle is a heavy neutrino, it must be of quasi-Dirac nature, since for heavy Majorana neutrinos $e^+ e^-$ and $e^\pm e^\pm$ events would be produced with the same rate. (Interestingly, a heavy quasi-Dirac neutrino is the most natural possibility to yield light neutrino masses $m_\nu \sim 1~\text{eV}$ with heavy mediators at the TeV scale, within an inverse seesaw mechanism~\cite{Mohapatra:1986bd}.) A new charged lepton is also possible~\cite{Dobrescu:2014esa} but we will not consider it, as it is not related to light neutrino mass generation (and the results are qualitatively very similar to the case of a heavy neutral lepton). Another possible interpretation is in terms of leptoquarks, and the CMS Collaboration has actually found a $2.4\sigma$ excess in searches for leptoquarks in the $eejj$ final state~\cite{CMS:2014qpa}, although the kinematical distributions are claimed not to be consistent with that hypothesis.

The cross section of the excess, $\sigma \sim 0.75~\text{fb}$ at an invariant mass around 2 TeV, also suggests the presence of new gauge interactions. In the minimal type-I seesaw the stringent limits on heavy neutrino mixing~\cite{delAguila:2008pw} lead to tiny production cross sections~\cite{Datta:1993nm}. In inverse type-III seesaw, the cross sections at this mass scale are a factor of 20 smaller~\cite{delAguila:2008hw}. Thus, new charged ($W'$) or neutral ($Z'$) interactions seem to be involved in resonant heavy neutrino production. The former scenario corresponds to the model and process used as benchmark in the CMS analysis, although any new $W'$ boson coupling to quarks and to $\ell N$, and not coupling to $\ell \nu$, would give a similar signal. In the latter case, the new $Z'$ boson must couple to quarks and not to the SM leptons so that $Z' \to e^+ e^-, \mu^+ \mu^-$ are absent and the stringent limits from these searches~\cite{Chatrchyan:2012oaa,Aad:2014cka} do not apply. On the other hand, process (\ref{ec:Zprod}) requires the $Z'$ boson to couple to $N$. An example of leptophobic $Z'$ boson coupling to heavy neutrinos is given by $\text{E}_6$ models~\cite{delAguila:2007ua}, and we will use this benchmark for our study. The interaction Lagrangian involved in the production is
\begin{equation}
\mathcal{L} = -g' Q \bar N \gamma^\mu N Z_{\mu}^{\prime} \,,
\label{eq:Lcn}
\end{equation}
with $Q=3/(2 \sqrt{6})$. The charged-current heavy neutrino decay takes place through the small mixing with LH leptons, {\it i.e.} the second term in Eq.~(\ref{ec:cc}). Additional decay modes $N \to Z \nu$, $N \to H \nu$ are also present (see~\cite{delAguila:2008cj} for a review).

For the sake of simplicity of our analysis, we have considered the case of only one heavy neutrino $N$ that can be produced through process~(\ref{ec:Wprod}) or (\ref{ec:Zprod}), and can mix with the three SM charged leptons. In the case of the $W'$ boson, the cross sections for different lepton flavors depend on $V_{\ell N} \equiv V_{\ell N}^R$ (remember that these mixings satisfy the unitarity constraint), whereas for the $Z'$ only the normalized quantities $V_{\ell N} \equiv V_{\ell N}^L/\sqrt{|V_{e N}^L|^2+|V_{\mu N}^L|^2+|V_{\tau N}^L|^2}$ are relevant, since the dependence on the sum of moduli squared mixings cancels with the $N$ width in the propagator. We have performed a fast simulation analysis using the leading-order (LO) generator {\sc Triada}~\cite{AguilarSaavedra:2009ik} for the signal, {\sc Pythia} 6.4~\cite{Sjostrand:2006za} for hadronisation and {\sc PGS 4}~\cite{pgs4} for the simulation of a generic LHC detector. The LO $W'$ and $Z'$ cross sections are scaled to next-to-leading order predictions by factors $k = 1.15$~\cite{Duffty:2012rf} and $k = 1.3$~\cite{Gao:2010bb}, respectively. The efficiencies for electrons and muons are slightly tuned to the ones in the CMS search by simulating $W'$ samples for $M_{W'} = 2.5$ TeV, $m_N = 1.25$ TeV and applying the same selection cuts on charged leptons and jets in~\cite{Khachatryan:2014dka}. Then, for several $W'$ and $Z'$ masses close to the CMS excess region, namely $M \simeq 2~\text{TeV}$, we have simulated samples corresponding to all flavor combinations $\ell_i,\ell_k=e,\mu,\tau$, including all possible heavy-neutrino decays, {\it i.e.} $W^{\prime *} \to u \bar d,c\bar s,t\bar b$ in the $W'$ model, and $N \to \ell W, \nu Z, \nu H$ in the $Z'$ model. For the $W'$ ($Z'$) signal we take $m_N = 1$ TeV ($m_N = 0.5$ TeV).

We find the favored mixing of the heavy neutrino by performing a $\chi^2$ analysis analogous to the ones in~\cite{AguilarSaavedra:2012gf,Aguilar-Saavedra:2013twa}. We consider the signal events in four bins $1.8~\TeV \leq m_{\ell\ell jj} \leq 2.2~\TeV$ and $2.2~\TeV \leq m_{\ell\ell jj} \leq 4.0~\TeV$, with $\ell \ell = ee,\mu \mu$, and denote them collectively as `$X$'. For each sample $\ell_i \ell_k$ an overall efficiency $\epsilon_{X}^{\ell_i \ell_k}$ ---which includes the appropriate branching ratios--- is determined as the fraction of simulated events passing the CMS selection cuts, and falling in each of the four bins $X$. Then, the number of expected signal events in each bin $X$ is:
\begin{equation}
S_{X}=\left(\frac{g'}{g}\right)^2 \sigma_{{\rm tot}}\,L \sum_{\ell_i\ell_k} \epsilon_{X}^{\ell_i \ell_k}\,|V_{\ell_i N}|^2 |V_{\ell_k N}|^2 \,,
\label{eq:Nll}
\end{equation}
with $\sigma_\text{tot}$ the total $\ell N$ ($NN$) production cross section for $g'=g$ and $L$ the integrated luminosity, which in the present case is $L=19.7\,{\rm fb}^{-1}$. In the fit, the number of observed events $N$ and expected background $B$ are $N_{ee} = 14 (4)$, $B_{ee} = 4.1 (2.3)$, $N_{\mu \mu} = 6(3)$, $B_{\mu \mu} = 6.0 (2.1)$ for the lower (higher) mass bins~\cite{Khachatryan:2014dka}.

\begin{figure}[t]
\begin{center}
\begin{tabular}{c}
\hspace{-0.3cm}\epsfig{file=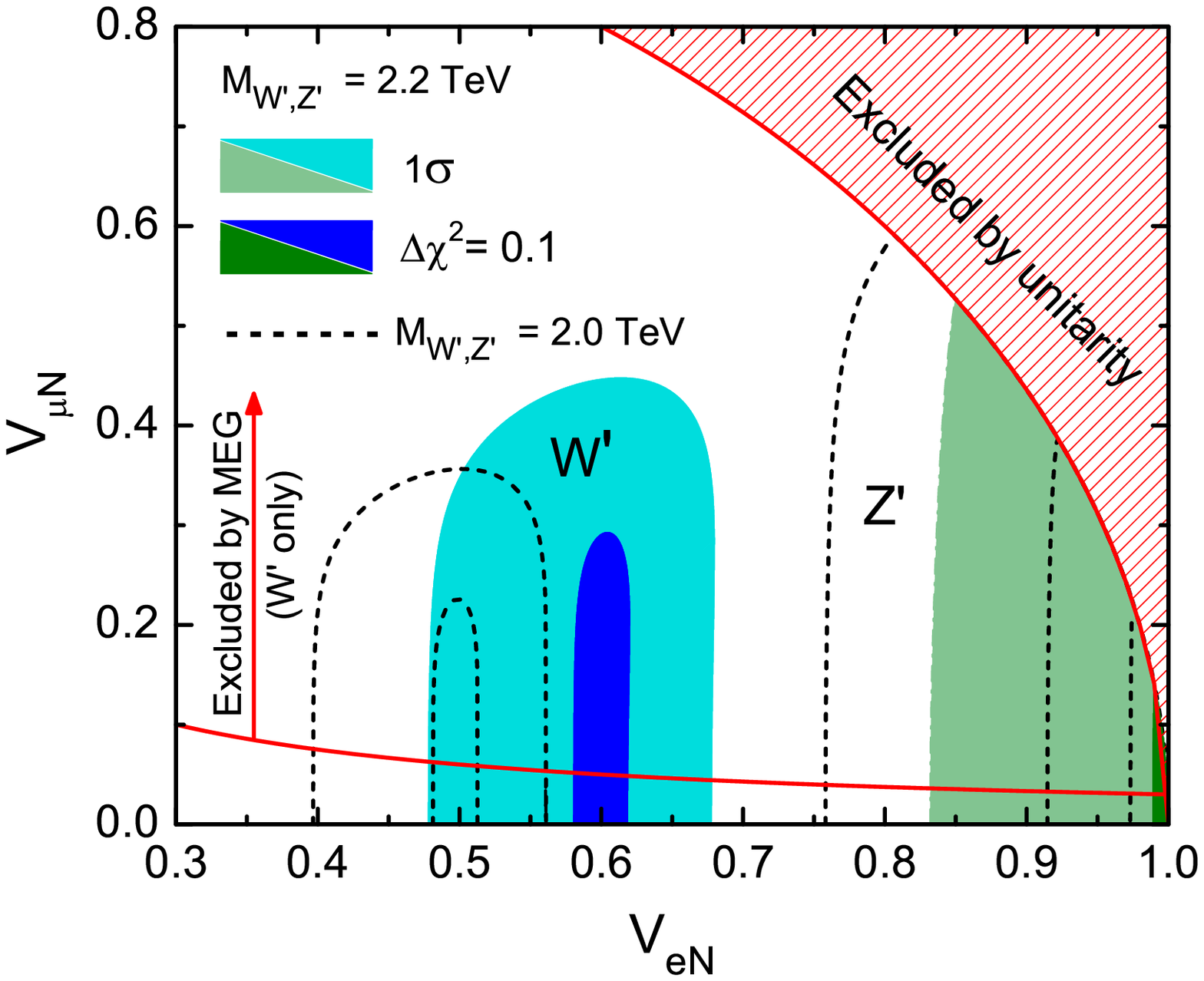,width=8.3cm,clip=} \\ 
\hspace{-0.2cm}\epsfig{file=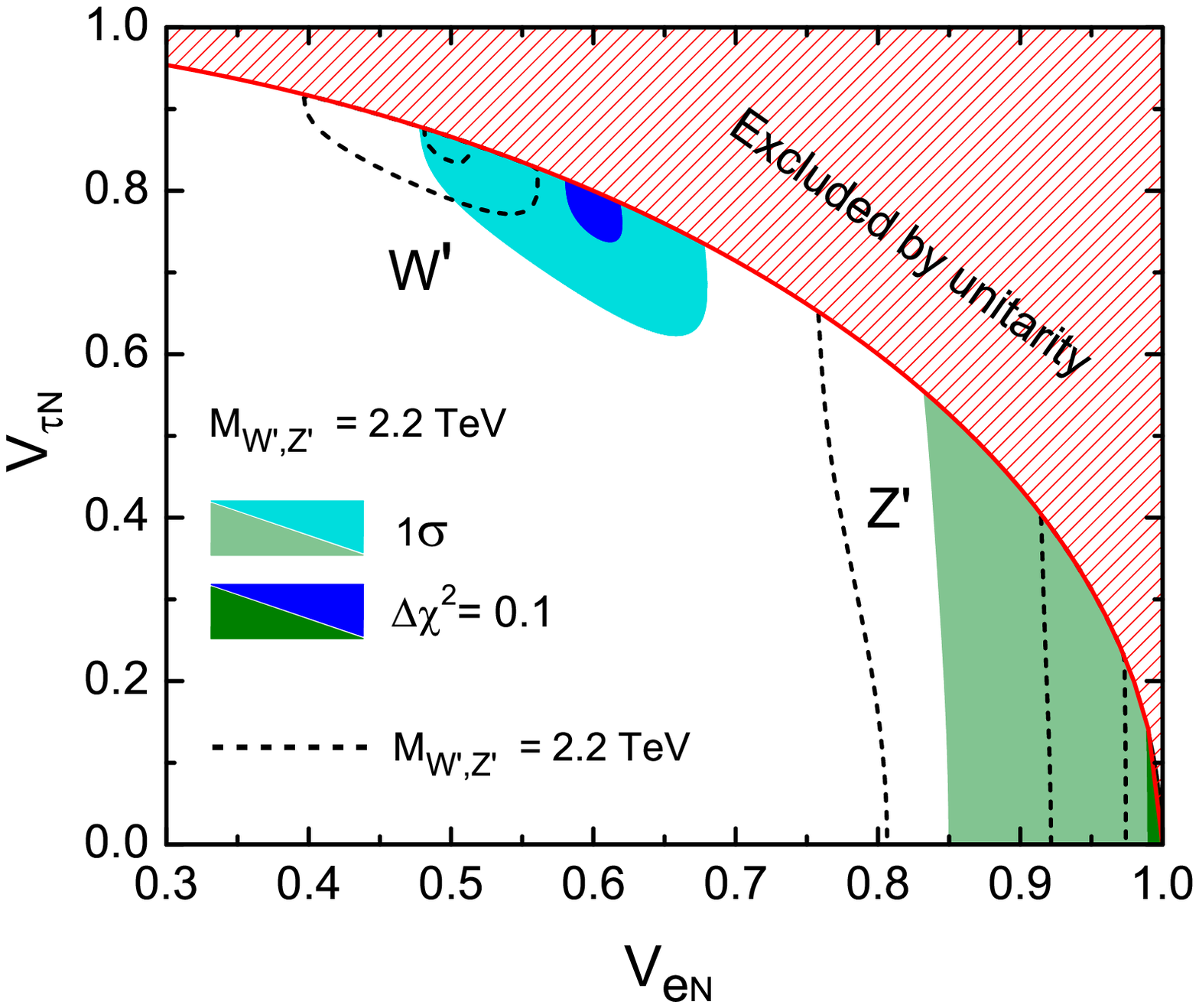,width=8.2cm,clip=}
\end{tabular}
\caption{Limits on heavy neutrino mixings at $1\sigma$ and with $\Delta \chi^2 < 0.1$ for $M = 2.2~\text{TeV}$ (light and dark green shaded regions, respectively) and $M = 2~\text{TeV}$ (dashed-line delimited regions) in the $(V_{eN},V_{\mu N})$ (top) and $(V_{e N},v_{\tau N})$ (bottom) planes. In the top panel we display the limit from $\mu \to e \gamma$ on the mixing in the $W'$ model~\cite{Cirigliano:2004mv} for illustration.}
\label{fig:mixplan}
\end{center}
\end{figure}

We present in Fig.~\ref{fig:mixplan} the fit results for two masses $M=2.2~\text{TeV},2.0~\text{TeV}$ and $g'=g$, which is allowed by dijet measurements~\cite{Aad:2014aqa}, $W' \to t \bar b$~\cite{Aad:2014xra} and $t \bar t$~\cite{Chatrchyan:2013lca} resonance searches for these masses. In all cases the variation of the $\chi^2$ around the minimum is rather mild so, instead of displaying the best-fit points, we show a region with $\chi^2 < 0.1$. For both the $W'$ and $Z'$ hypotheses, the fit is better for $M = 2.2$ TeV. For the $W'$ case the size of the signal can be accommodated with a non-zero mixing with the $\tau$ lepton which, as we will see below, smears the kinematical distributions. Whereas, for the $Z'$ the mixing with the electron must be dominant since the signal is smaller.
Notice also that while, in principle, a considerable $\mu N$ mixing is allowed by the fit to CMS data, the conservative bound coming from $\text{Br}(\mu \to e \gamma)$~\cite{Adam:2011ch} constrains $V_{\mu N}$ to be small in the case of the $W'$. (In the $Z'$ case the cross section is independent of the absolute normalisation of the mixings, so the $\mu \to e\gamma$ constraints can be evaded for small enough $V_{\ell N}^R$.) The bounds on $\tau \to e \gamma$ are fulfilled for any values of the mixing parameters.

\begin{figure}[htb]
\begin{center}
\epsfig{file=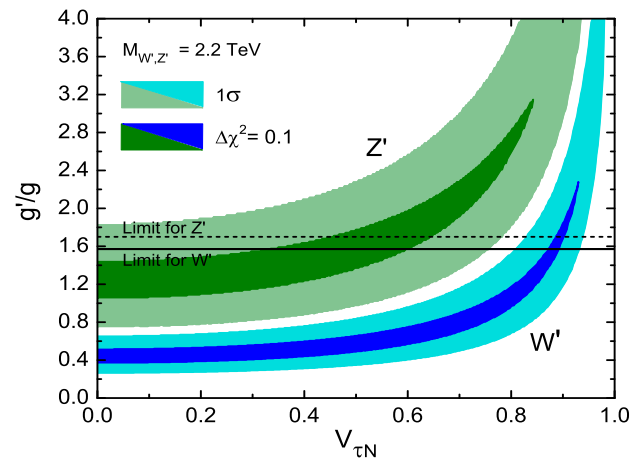,width=8.2cm,height=6cm,clip=} 
\caption{Same as in Fig.~\ref{fig:mixplan} but in the $(V_{\tau N},g'/g)$ plane. The solid (dashed) horizontal line corresponds to the upper limit on $g'/g$ that result from dijet and $t \bar b$ (dijet and $t \bar t$ resonant) production in the $W'$ ($Z'$) case.}
\label{fig:VtaugRL}
\end{center}
\end{figure}

Provided $V_{\mu N}$ is small, the size of the dielectron signal is controlled by the ratio $g'/g$ and the mixing $V_{\tau N}$. We present in Fig.~\ref{fig:VtaugRL} the allowed regions in the $(V_{\tau N},g'/g)$ plane for $M = 2.2$ TeV and $V_{\mu N} = 0$, together with the 95\% confidence level upper limits on $g'/g$ that result from dijet and $t \bar b$ production in the case of the $W'$ (solid horizontal line), and from dijet and $t \bar t$ resonant production for $Z'$ (dashed horizontal line). We observe that there is enough room to fit the size of the dielectron excess with couplings $g' \sim g$ and various sizes of $V_{\tau N}$, including $V_{\tau N} = 0$. 

\begin{figure*}[t]
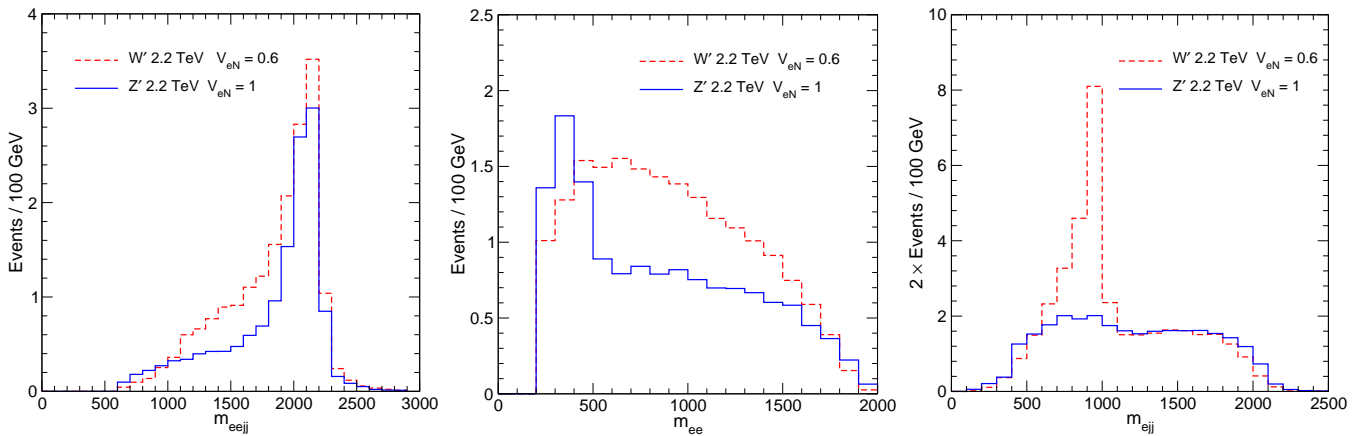

\begin{center}
\begin{tabular}{ccc}
\epsfig{file=fig3a,height=5.7cm,clip=} & 
\epsfig{file=fig3b,height=5.7cm,clip=} & 
\epsfig{file=fig3c,height=5.7cm,clip=}
\end{tabular}
\caption{Invariant mass distributions for two selected benchmarks, taking $V_{\mu N} = 0$. Left: $eejj$ invariant mass. Center: dilepton invariant mass. Right: heavy neutrino reconstructed mass under the $W'$ production hypothesis.}
\label{fig:histgr}
\end{center}
\end{figure*}

Having shown that the size of the possible signals in the two $eejj$ invariant mass bins can be accommodated by $W'$ or $Z'$ production in various mixing scenarios, we turn our attention to differential distributions, which have motivated the CMS claim that the excess is not likely due to the process~(\ref{ec:Wprod}). We present the relevant distrubutions for the $W'$ and $Z'$ signals in Fig.~\ref{fig:histgr} for $M = 2.2$ TeV, $g'=g$, $V_{\mu N} = 0$ and $V_{eN} = 0.6 (1)$ for $W'$ ($Z'$). The first one to consider is the $eejj$ invariant mass (left), which obviously peaks around $2.2$ TeV for $W'$, as expected from the kinematics in~(\ref{ec:Wprod}). A small ``shoulder'' at lower $m_{eejj}$ is present due to the missing energy in decays involving one $\tau$ lepton. But, remarkably, $m_{eejj}$ also peaks around 2.2 TeV for the $Z'$, despite these events have two extra jets at the partonic level. The reason is that when defining $m_{eejj}$ the two jets with larger transverse momentum are chosen, making this quantity close to the $Z'$ reconstructed mass, which involves two more jets.
The dilepton invariant mass $m_{ee}$ distribution (central panel) seems to agree with the one observed in~\cite{Khachatryan:2014dka} for both the $W'$ and $Z'$ cases. We point out that, for $W'$, the enhancement at low $m_{ee}$ is produced by events involving a $\tau$ lepton; if $V_{\tau N} = 0$ the distribution has a maximum around $m_{ee} = 1$ TeV. (This distribution also depends on the heavy neutrino mass, and for larger $m_N$ it becomes flatter.) Finally, in Fig.~\ref{fig:histgr} (right) we present the heavy neutrino mass reconstructed under the hypothesis of $W'$ production~(\ref{ec:Wprod}). Since one does not know {\it a priori} which of the two electrons results from the $N$ decay, one can consider the invariant mass of the two combinations $m_{e_1 jj}$, $m_{e_2 jj}$, obtaining a plot with two entries per event, which for $W'$ production displays a peak at $m_N = 1$ TeV (the actual value used in our simulation) whereas it is flat for $Z'$ production. This distribution has not been made available in the CMS analysis, therefore it is difficult to conclude whether data prefers one or the other interpretation. However, we point out that for the $W'$ benchmark the heavy neutrino peak contains 10 events out of 18, so it is not at all obvious that with the available statistics it should necessary show up. In any case, for $Z'$ production there is no such peak.

To conclude, we remark that if the CMS excess is due to either of the $W'$ or $Z'$ production processes  discussed, trilepton signals should also show up with a moderate increase in the statistics and/or a dedicated search. In the case of $W'$, $1/3$ of the $N$ decays in (\ref{ec:Wprod}) involve a $t \bar b$ pair, which is a trademark for this process. With the statistics available in the LHC run 2, final states $ee t\bar b$ with reconstructed top quarks could be searched for. In the case of $Z'$, trilepton and four lepton signals appear when one or the two $W$ bosons in (\ref{ec:Zprod}) decay leptonically. In both scenarios, the predicted signals are compatible with small excesses found by the CMS Collaboration~\cite{Chatrchyan:2014aea}, but this should be confirmed or discarded with more statistics. Searches in the $e\mu$ final state are also interesting, as potential excesses may show up in this channel too. Conversely, the absence of a signal would further constrain the heavy neutrino coupling and mixing parameter space. In this respect, a classification of $\ell \ell jj$ events by missing energy is also useful~\cite{AguilarSaavedra:2012gf} to identify the secondary leptons from $\tau$ decays.

In summary, in this Letter we have addressed two possible interpretations of a CMS excess~\cite{Khachatryan:2014dka} in terms of a new $W'$ or $Z'$ boson and a heavy neutrino, in a general flavor mixing context. We stress that, from the theoretical point of view, there is no compelling argument in favor of heavy neutrinos mixing with only one charged lepton. On the contrary, in view of the large mixing observed in neutrino oscillations, non-zero mixing with the three lepton generations is somehow expected. Actually, the interpretation of the excess in terms of a $W'$ boson and a heavy neutrino is more consistent with data within this more general framework. On the other hand, even if the $W'$ interpretation proves to be incorrect ---due to the absence of enhancements in certain kinematical distributions--- the $Z'$ hypothesis explains the CMS observation without producing other kinematical peaks apart from the $m_{eejj}$ one already observed.
\\

{\bf Acknowledgements:} this work has been supported by FCT (Portugal) through the projects  EXPL/FIS-NUC/0460/2013, CERN/FP/123580/2011 and PEst-OE-FIS-UI0777-2013, by MICINN (Spain) project FPA2010-17915 and by
Junta de Andaluc\'{\i}a projects FQM 101 and FQM 6552. F.R.J. also thanks the CERN Theory Division for hospitality.

\end{document}